\documentclass[twocolumn,showpacs,preprintnumbers,amsmath,amssymb,prl]{revtex4}

\usepackage{graphicx}
\usepackage{bm}

\begin{document}

\title{"Double swallow-tail" singularity and glass-glass transition in
a quasibinary system.}

\author{V. N. Ryzhov}
\affiliation{Institute for High Pressure Physics, Russian Academy
of Sciences, Troitsk 142190, Moscow Region, Russia}
\affiliation{Moscow Institute of Physics and Technology, 141700
Moscow, Russia}

\author{E. E. Tareyeva}
\affiliation{Institute for High Pressure Physics, Russian Academy
of Sciences, Troitsk 142190, Moscow Region, Russia}

\date{\today}

\begin{abstract}
The system with the square shoulder (SS) potential is considered
in the frame of Mode Coupling Theory (MCT) approach. An
approximation for the structure factor is used that emphasizes the
quasibinary character of the system. The qualitative phase diagram
is constructed that includes continuous and discontinuous
glass--glass transitions. The phase diagram is governed by two
swallow tails connected with two $A_4$ singularities.
\end{abstract}

 \maketitle

A large number of papers dedicated to the liquid-glass transition
have been  published during last decades. They include results of
real experiments, computer simulations and different theoretical
approaches. Nowaday the most popular and the most cited of the
various theories of glasses are based on mean--field replica
approach \cite{ParZamp} and the so called Random First Order
Transition theory (RFOT) \cite{Wol}, both based on analogies with
the well-developed equilibrium statistical mechanics of spin
glasses. Numerous results were obtained in the framework of the
so-called mode-coupling theory (MCT) (see, e.g., the pioneering
work \cite{1}, the review \cite{2}, and also the detailed
presentation in the recent monograph \cite{4}). Although to-day
the investigations of dynamical heterogeneities (see e.g.
\cite{heter}) in glassy systems make favourable the different
variants of overrunning MCT, one can affirm that it was for a long
time the only consistent theory describing details of the
transitions in supercooled liquids. Despite suffering of a shift
of the actual glass transition value, MCT provides a good
description of the experimental data. In the present paper we
remain in the frame of the traditional MCT.

In the framework of MCT, we propose the qualitative description of
liquid-glass and glass-glass transitions in a system of hard
spheres with an additional repulsive step in the potential
(square-shoulder (SS) system). The MCT equations demonstrate the
$A_4$ bifurcation singularities and $A_3$ end points. In the
framework of the approximation \cite{TMF}, corresponding to the
quasibinary nature of the SS system \cite{14}, we show that there
is some symmetry of the MCT equations, which leads to the
duplication of singularities. As far as we know, the double
swallow tail is described in the present paper in the first time.

In MCT, the system dynamics is  described in terms of the
autocorrelation function of the density fluctuations
 $\Phi_q(t)= <\rho _q(t)\rho_{-q}(0)>/<\rho _q(0)\rho_{-q}(0)>$
 well known in the theory of liquids
(see, e.g., Ref.\cite{3}). Here, $\rho _q(t)$ is the Fourier
transform of the system density. The autocorrelation  function
satisfies the equation
\begin{eqnarray}
\frac{\partial ^2 \Phi_q(t) }{\partial t^2} &+& \nu _q
\frac{\partial \Phi_q(t)}{\partial t} +\Omega_q^2 \Phi_q(t) + \nonumber\\
&+&\Omega^2_q \int_0^t dt'
m_q(t-t')\frac{\partial\Phi_q(t')}{\partial t'} =0, \label{1}
\end{eqnarray}
where $\nu _q$ is white noise and
 $\Omega_q$ the characteristic frequency. The memory function
$m_q(t)$ has the form
\begin{equation}
m_q(t)=\frac{1}{2} \int \frac {d^3 k}{(2\pi )^3} V_{\vec q,\vec
k}\Phi_{\vec k}(t) \Phi_{\vec q-\vec k}(t) \label{2}
\end{equation}
The interaction potential of the system particles is included in
the vertex function
\begin{equation}
V_{q,k}=\rho S_{\vec q}S_{\vec k}S_{\vec q-\vec k}[\vec q \vec k
c_k + \vec q (\vec q - \vec k) c_{\vec q - \vec k}]^2 /q^4
\label{3}
\end{equation}
through the static structure factor $S_q$ of liquid and the direct
correlation function $c_q$ \cite{3}. These two quantities are
related:
\begin{equation}
S_q=1/(1-\rho c_q). \label{4}
\end{equation}
The behavior of the solution of Eq.~(\ref{1}) at large times
determines relaxation processes in the system \cite{1,2,4}. As $t
\to \infty$, the algebraic equation for the limit correlation
function $f_q=\Phi_q(\infty)$ :
\begin{equation}
\frac{f_q}{1-f_q}=\frac{1}{2} \int \frac {d^3 k}{(2\pi )^3}
V_{\vec q,\vec k} f_{\vec k} f_{\vec q - \vec k} \label{5}
\end{equation}
can be obtained from Eq.~(\ref{1}). Eq.~(\ref{5}) always has the
trivial solution $f_q=0$ corresponding to the liquid (ergodic)
phase. Eq.~(\ref{5})  can also have a nonzero solution $f_q>0$
corresponding to a nonergodic glass phase. If $f_q=0$ is a
bifurcation point of Eq.~(\ref{5}) the liquid-glass transition is
continuous, if no, then $f_q$  changes jumpwise from zero to a
certain value $f_q^c$ at the transition point. Hence, the value
$f_q$ can be regarded as an order  parameter (or a nonergodicity
parameter) for the liquid-glass transition.

The MCT was first used to describe the transition to the glass
state in the system with the hard-core potential in \cite{ParZamp}
and then in a large number of various systems. Its applicability
was confirmed experimentally (see \cite{2,4,5}). At the same time,
papers where the possibility of describing the glass-glass
transition for certain potentials in the MCT framework have
recently appeared: in this case, the glass characterized by the
nonergodicity parameter $f_q^{(1)}$  transforms jumpwise into
another glass with the nonergodicity parameter $f_q^{(2)}$. Such a
transition was predicted for systems with a potential consisting
of a solid core and a very narrow and deep attractive well in
\cite{6,7,8,9}. In this case, the first glass state is determined
by repulsion as for the system of hard spheres, while the second
glass state is determined by attraction. It is the competition
between these two states that determines the glass-glass
transition. The glass-glass transition line continues the
liquid-glass transition line smoothly to the glass region and ends
at a third-order bifurcation point.

Now we consider a system in which the  attracting well is changed
for a repulsive step - SS system. This system and its smoothed
versions were widely studied
\cite{TMF,14,10,11,12,13,15,16,17,18,19,20,sperl,sperl1,das}. It
was shown that a series of unusual phenomena are observed in the
system, including maxima on the melting curve and structural
transitions in the solid phase. In addition, thermodynamic
anomalies like ones observed in water were discovered in the
system with a smoothed version of the potential.

The potential of the system under consideration has the form
\begin{equation}
\Phi (r)=\left\{
\begin{array}{lll}
\infty , & r\leq d \\
\varepsilon , & d <r\leq \sigma  \\
0, & r>\sigma%
\end{array}%
\right.  \label{6}
\end{equation}
where $d$ is the diameter of the hard core and $\sigma $ and
$\epsilon $ are the diameter and height of the repulsive step. In
the papers \cite{sperl,sperl1,das} the MCT equations obtained for
this system using different approximations for the static
structure factor were solved numerically and slightly different
results were obtained, drastically depending on the approximation.

It can be seen that all the variants of the structure factor used
in the mentioned papers have an obvious common feature: the first
peak, corresponding to the diameter of the repulsive step,
decreases with the increase of the density, and the second peak,
determined by the hard core, increases. In this paper, we try to
demonstrate the main features of this quasibinary system in the
frame of a very simple approximation for the structure factor
$S(q)$ \cite{TMF}:
\begin{equation}
S(q) \approx 1+ A \delta (q-k_1) +B \delta (q-k_2), \label{8}
\end{equation}
where $k_1$ and $k_2$ are the positions of the first and second
maxima of the structure factor. The parameters $A$ and $B$ are
determined by the area under the corresponding maxima \cite{TMF}.
As it was shown in Ref.~\cite{TMF}, this approximation works well
for low densities and gives qualitatively correct result for the
high density region. Our aim is to obtain a schematic phase
diagram in terms of three control parameters
\begin{equation}
x=\frac{S(k_2)}{S(k_1)}\frac{k_2^2}{k_1^2}, 
\label{Eq/Pg4/1:swal_2}
\end{equation}

\begin{equation}
a=\frac{S(k_1)k_1}{8 \pi^2 \rho }, 
\label{Eq/Pg4/2:swal_2}
\end{equation}

\begin{equation}
b=\frac{S(k_2)k_2}{8 \pi^2 \rho }. 
\label{Eq/Pg5/1:swal_2}
\end{equation}

Substituting approximation (\ref{8}) in Eq.~(\ref{5}), we obtain
an approximate system of equations for the nonergodicity
parameters $f(k_1)$ and $f(k_2)$:
\begin{eqnarray}
\frac{f(k_1)}{1-f(k_1)} &=&  \frac {S(k_1)k_1}{8\pi ^2 \rho }
\left(Af(k_1) + \frac{k_2}{k_1}Bf(k_2)\right)^2 \nonumber\\
\frac{f(k_2)}{1-f(k_2)} &=&  \frac {S(k_2)k_2^2}{8\pi ^2 k_1 \rho
} \left(Af(k_1) + \frac{k_2}{k_1}Bf(k_2)\right)^2 \label{9}
\end{eqnarray}

The system can be considered as a mixture of hard spheres with
diameters $d$ and $\sigma $ \cite{14}. It can be seen \cite{TMF},
that the glass consisting of particles with the larger diameter
$\sigma $ exists for small densities, while the glass consisting
of hard spheres with the diameter $d$ prevails for high densities.
If these two types of local glass structure are taken into
account, then the question arises, whether the glass-glass
transition between these states exists. Indeed, an analysis of
Eqs.~(\ref{9}) shows that there is a fourth-order singularity
point $A_4$ for this system. This singularity corresponds to the
division of the space of three control parameters into three parts
corresponding to the absence of nontrivial solution of the
equations, to the part with two solutions, and the remaining part
with four solutions, by the surface of the figure called "swallow
tail" (see \cite{Arn1,Arn2,osob}).

Our approximation emphasizes the quasibinary nature of the system
and enables one to obtain a qualitative phase diagram of the
system. The approximation leads to the new kind of the symmetry of
the equations and to a degeneracy, both having a physical ground.
The duplication of the singularities demonstrates one of the
consequences of the degeneracy - we obtain "double swallow tail".

Despite a lot of works (see, for example,
\cite{4,GoSp02,flach,sperl}) where the higher order singularities
in the MCT were discussed, the degenerate cases and the
multiplication of singularities were not considered in the
literature.

Let us note that although our results are of qualitative
character, the obtained picture can be considered in as a scenario
for "the ideal glass-glass transition" in quasibinary systems.

We now pass directly to the problem of finding the bifurcation
points of Eqs.~(\ref{9}). As is well known (see, e.g. \cite{vain})
the uniqueness of solution of two functional equations
$$\chi _1(z_1,z_2) = 0; \hspace{3mm} \chi _2(z_1,z_2) = 0 $$
fails at the points where the determinant of derivarives $||\det
\frac{\partial\chi _i}{\partial z_k}||$ is zero.

Let us use the standard MCT notations (see \cite{4}) and rewrite
the system (\ref{9}) in the following form:
\begin{eqnarray}
\frac{f_1}{1-f_1} &=& (a f_1+b f_2)^2  [=F_1], \nonumber\\
\frac{f_2}{1-f_2} &=& x (a f_1+b f_2)^2  [=F_2],
\label{Eq/Pg1/1:swal}
\end{eqnarray}
where it is easy to see that these equations are invariant under
the transformation

$$f_1 \to f'_2; f_2 \to f'_1$$
$$a \to \sqrt{x'}b'; b \to \sqrt{x'}a'; x \to 1/{x'}$$

This means that for every solution of (\ref{Eq/Pg1/1:swal}) there
will be another one and the same can be said about every point
$A_k$.

At the bifurcation points we have $||\det [\delta _{qk}-A_{qk}]||
= 0$, where
\begin{eqnarray} A_{11}&=&(1-f_1^c)^2 F_{1,1}',
\nonumber\\
A_{12}&=&(1-f_1^c)(1-f_2^c) F_{1,2}',\nonumber\\
A_{21}&=&(1-f_1^c)(1-f_2^c) F_{2,1}',\nonumber\\
A_{22}&=&(1-f_2^c)^2 F_{2,2}'.\nonumber\\
\end{eqnarray}

Here $F_{1,1}'=2ay$;  $F_{1,2}'=2by$; $F_{2,1}'=2axy$;
$F_{2,2}'=2bxy$.

The determinant $||\det A_{q,k}||=0$, and this fact is the
consequence of the mentioned degeneracy. In fact, the matrix
$A_{qk}$ eigenvalues in our case are:
$$\lambda _1=0, \lambda _2=A_{11}+A_{22}$$
and the bifurcation condition can be written in the form
\begin{equation}
(1-f_1^c)^2 2ay+(1-f_2^c)^2 2bxy =1 \label{Eq/Pg1/2:swal}
\end{equation}

To find the points of bifurcation of the system it is technically
convenient to use the new variable defined as follows:
$$y=a f_1+b f_2.$$

The equations ( \ref{Eq/Pg1/1:swal}) can be rewritten in terms of
$y$. Now
$$f_1=\frac{y^2}{1+y^2}; \hspace{4mm} f_2=\frac{xy^2}{1+xy^2}$$
so that
\begin{equation}
y=\frac{ay^2}{1+y^2}+\frac{bxy^2}{1+xy^2} \label{Eq/Pg1/3:swal}
\end{equation}

The derivative of this equation relative to $y$ gives
\begin{equation}
1=\frac{2ay}{(1+y^2)^2} +\frac{2bxy}{(1+xy^2)^2}
\label{Eq/Pg2/1:swal}
\end{equation}

This equation coincides with the bifurcation condition
(\ref{Eq/Pg1/2:swal}), if one takes into account that
\begin{eqnarray}
1-f_1&=&\frac{1}{1+y^2}, \nonumber\\
1-f_2&=&\frac{1}{1+xy^2}. \nonumber
\end{eqnarray}
Now the problem of obtaining the bifurcation points of initial
equations reduces to the system of equations
(\ref{Eq/Pg1/3:swal})-(\ref{Eq/Pg2/1:swal}), that is the equations
$$\tilde \Psi (y)=0; \hspace{4mm} \tilde \Psi' _y =0$$
with
$$\tilde \Psi (y) =
y-\frac{ay^2}{1+y^2}+\frac{bxy^2}{1+xy^2}.$$

Let us make two remarks. First, it is easy to see that the point
$y=0$ is not a bifurcation point of our equations, so that one can
use $\frac{\tilde \Psi (y)}{y}$ instead of $\tilde \Psi (y)$ for
our purpose. Second, as far as the bifurcation properties are
concerned it is possible to consider only the numerator as the
main functional. All the equations for singularities (including
higher orders) for these functionals will differ only by additive
terms proportional to the lower order derivatives, which are zero.
So, we can use the following functional:
\begin{equation}
\Psi (y,a,b,x)= ay(1+xy^2)+bxy(1+y^2)-(1+y^2)(1+xy^2).
\label{Eq/Pg2/2:swal}
\end{equation}
We have for the $A_2$ singularities:
\begin{eqnarray}
&&\Psi (y,a,b,x)=0;\nonumber \\
&&\Psi'_y(y,a,b,x)=0; \label{Eq/Pg2/3:swal}
\end{eqnarray}
for the $A_3$ singularities:
\begin{eqnarray}
&&\Psi (y,a,b,x)=0; \nonumber \\
&&\Psi'_y(y,a,b,x)=0; \nonumber \\
&&\Psi''_y (y,a,b,x)(y)=0; \label{Eq/Pg2/4:swal}
\end{eqnarray}
and for the $A_4$ singularity:
\begin{eqnarray}
&&\Psi (y,a,b,x)=0; \nonumber \\
&&\Psi'_y(y,a,b,x)=0; \nonumber \\
&&\Psi''_y (y,a,b,x)(y)=0; \nonumber \\
&&\Psi'''_y (y,a,b,x)(y)=0; \label{Eq/Pg2/5:swal}
\end{eqnarray}
while $\Psi_y^{IV}(y,a,b,x)\neq0$.

The explicit forms for these derivatives are:

\begin{eqnarray}
\Psi'_y(y,a,b,x)(y)&=& a(1+3x y^2)+bx(1+3y^2)-\nonumber \\
&-&2y(1+x+2xy^2), \label{Eq/Pg3/1:swal}
\end{eqnarray}

\begin{equation}
\Psi''_y(y,a,b,x)(y)=6xy(a+b)-2-2x-12xy^2, \label{Eq/Pg3/2:swal}
\end{equation}

\begin{equation}
\Psi'''_y(y,a,b,x)(y)=6x(a+b)-24xy, \label{Eq/Pg3/3:swal}
\end{equation}
and $\Psi_y^{IV}(y,a,b,x)=-24x$.

Technically it is convenient to proceed in the following way. The
equations (\ref{Eq/Pg2/3:swal}) with  (\ref{Eq/Pg2/2:swal}) and
(\ref{Eq/Pg3/1:swal}) are two linear equations for the variables
$a$ and $b$ which can be easily solved:
\begin{eqnarray}
a&=&\frac{(1-xy^2)(1+y^2)^2}{2y^3(1-x)};\nonumber \\
b&=&-\frac{(1-y^2)(1+xy^2)^2}{2xy^3(1-x)}. \label{Eq/Pg3/4:swal}
\end{eqnarray}

Substituting of these $a$ and $b$  in (\ref{Eq/Pg3/2:swal}) and
(\ref{Eq/Pg3/3:swal}) gives;

\begin{equation}
\Psi''\sim \mu _2=-6+2y^2(1+x)-6xy^4 \label{Eq/Pg3/5:swal}
\end{equation}
and
\begin{equation}
\Psi''' \sim \mu _3=-1+(1+x)y^2-5xy^4 \label{Eq/Pg3/6:swal}
\end{equation}

The points $\{a,b\}$ (at  fixed values of $x$) for which there are
some real solutions $y>0$ satisfying both of the equations
(\ref{Eq/Pg3/4:swal}) make the lines of points $A_2$. In fact, at
the fixed value of $x$ and the set of $a$ the first of the
equations (\ref{Eq/Pg3/4:swal}) was solved. The obtained value of
$y$ at the fixed $x$ then give value of $b$ through the second
equation (\ref{Eq/Pg3/4:swal}). Those points ($a,b,x$) for which
in addition to the existence of $y$, the values $x$ and $y$ are
related through (\ref{Eq/Pg3/5:swal}), present the line of $A_3$
singularity.

\begin{figure}
\includegraphics[width=8cm]{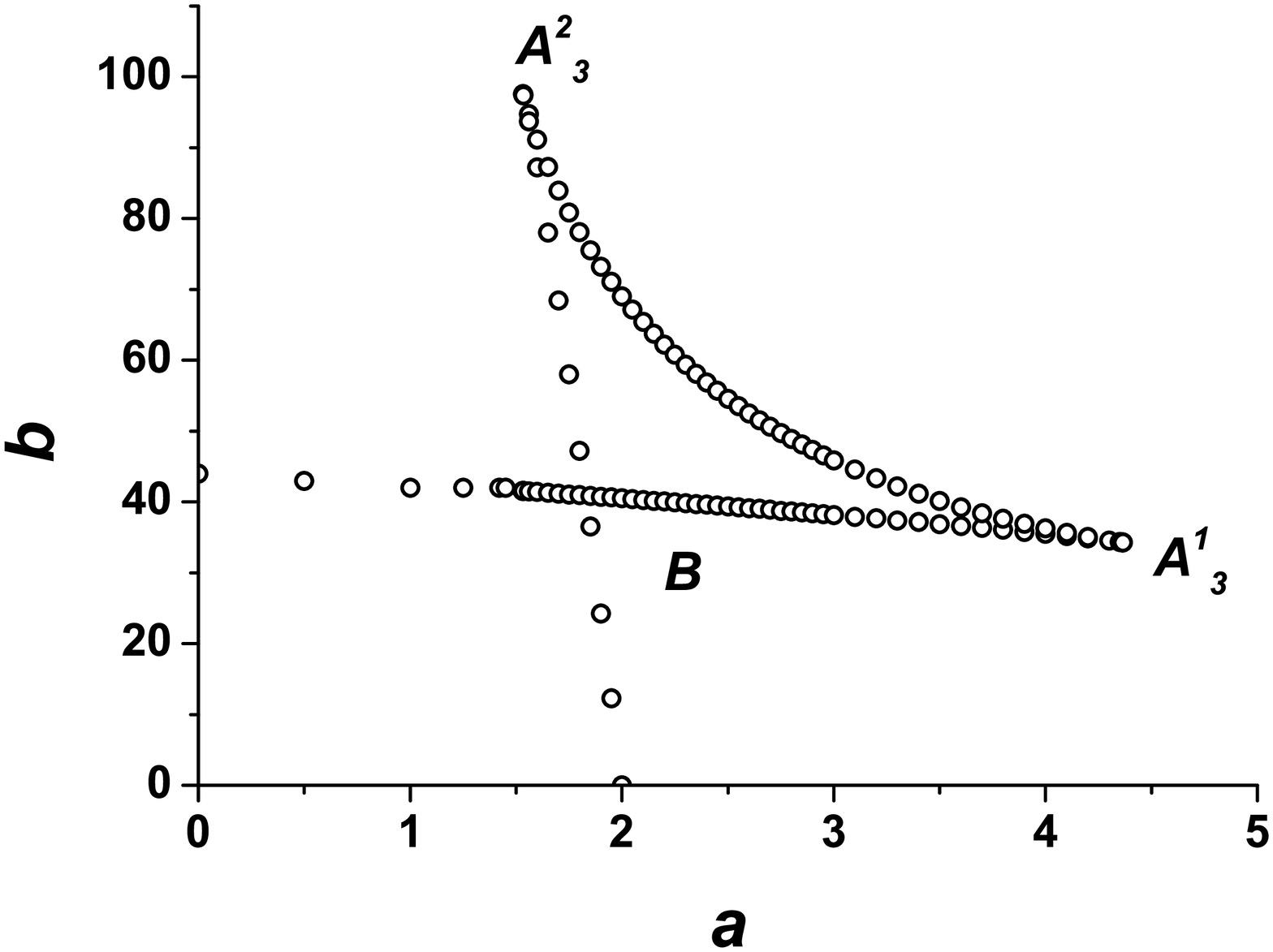}

\caption{\label{fig:fig1} Swallow-tail cross section by the plane
$x=0.002$.}
\end{figure}

It is easy to solve the system of equations $\mu _2=0$ and $\mu
_3=0$ and to obtain two solutions for the $A_4$ singularity:
$x_{01}=0.0294373, y_{01}=2.41421$ and $x_{02}=33.9706,
y_{02}=0.4142$.The corresponding values of $a$ and $b$ are
determined by (\ref{Eq/Pg3/4:swal}): $a_{01}=1.41421;
b_{01}=8.24264$ and $a_{02}=1.41421, b_{02}=0.24264$ with
 $b_{01} \sqrt{x_{01}} = b_{02}\sqrt{x_{02}}$.

\begin{figure}
\includegraphics[width=8cm]{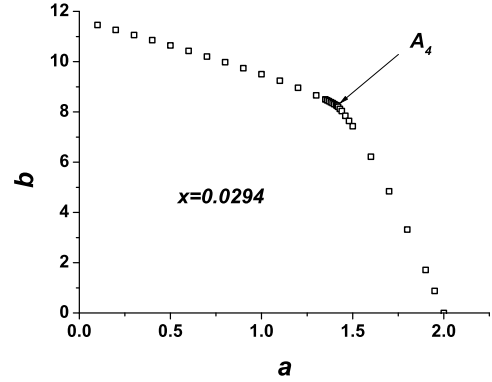}

\caption{\label{fig:fig2} Swallow-tail cross section by the plane
$x=x_{01}=0.0294$ - the $A_4$ singularity point.}
\end{figure}

In Fig.~\ref{fig:fig1} the lines of $A_2$ points at $x=0.002 $ are
presented, as well as two $A_3$ points and the crossing point $B$.
The similar pictures can be obtained for any value of $x$ up to
$x_0$ corresponding to the $A_4$ singularity. The section at
$x_{01}$ is shown in Fig.~\ref{fig:fig2}. The sections for $x$
between $x_{01}$ and $x_{02}$ present smooth curves. At $x=x_{02}$
the figure is similar to Fig.~\ref{fig:fig2} and for $x>x_{02}$ -
similar to Fig.~\ref{fig:fig1}. For given values of $x$ and $y$ it
is easy to obtain  $f_1$ and $f_2$.

Taking into account the maximum principle (see \cite{4}) for the
functions $f_1$ and $f_2$, one can show that the transition lines
are $B A_3^{1}$ for $x<x_{01}$ and $B A_3^{2}$ for $x>x_{02}$ (see
Fig.~\ref{fig:fig1}). These lines are related to jumpwise
glass-glass transitions. So, due to symmetry property we obtain
two swallow tails for the surfaces of $A_2$ singularities and two
$A_4$ singularities. Each swallow tail separates one from another
three parts of the 3D space of the variables $x,a,b$. In one
(curved pyramid) there are 4 solutions of the initial equation
$\Psi (y,a,b,x)=0$ ; in the neighboring part there are 2 solutions
and in the remainder - no solution. The number of positive
solutions, we are interested in, may be different from the
mentioned. This remaining part (for example, for $x=0.002$ it
corresponds approximately to $b<40$ and $a<2$) has the only
solution $y=0$, which satisfies the Eq. $\tilde\Psi=0$. This gives
$f_1=0$ and $f_2=0$ - the ergodic (liquid) phase. If we take into
account the maximum principle for the bifurcated functions
\cite{4}, we have to discard some parts of the swallow tails as a
phase boundary: the almost horizontal line for the first
swallow-tail singularity and almost vertical line for the second
one.

\begin{figure}
\includegraphics[width=8cm]{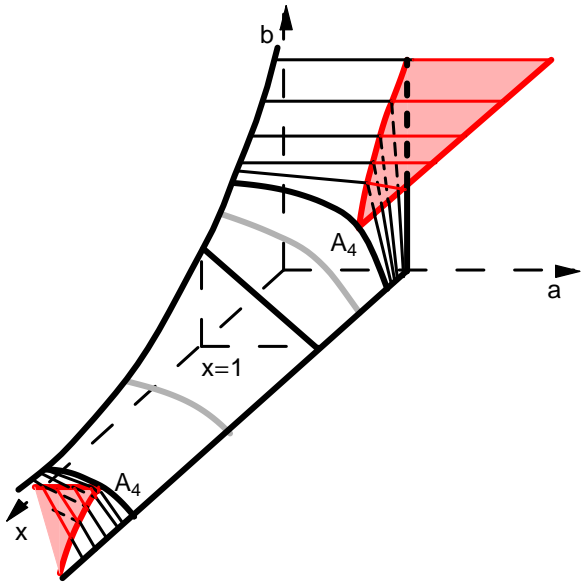}

\caption{\label{fig:fig3} (Color online) The total phase diagram.}
\end{figure}

\begin{figure}
\includegraphics[width=8cm]{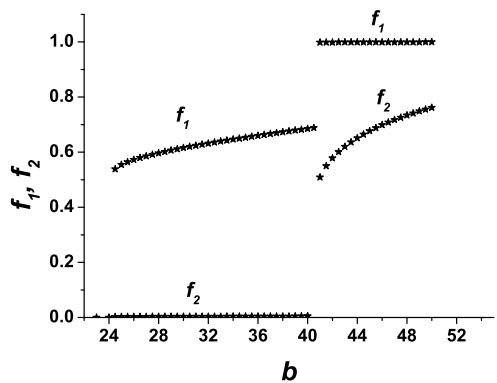}

\caption{\label{fig:fig4} $f_1$ and $f_2$ as functions of $b$ for
$x=0.002$ and $a=1.9$.}
\end{figure}

\begin{figure}
\includegraphics[width=8cm]{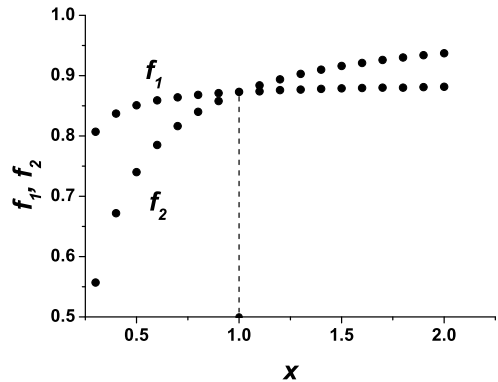}

\caption{\label{fig:fig5} $f_1$ and $f_2$ as functions of $x$ for
$a=1.5$ and $b=1.5$.}
\end{figure}

The resulting phase diagram of our system in the $3D$ space of the
variables $x,a,b$ (\ref{Eq/Pg4/1:swal_2})-(\ref{Eq/Pg5/1:swal_2})
is shown schematically in the Fig.~\ref{fig:fig3}. The space
inside the tube corresponds to the ergodic liquid phase ($f_1=0,
f_2=0$). The liquid-glass transition takes place at the tube
surface and corresponds to the jumps of the functions $f_1$ and
$f_2$ (from zero). Outside the tube we have glass states. The tube
surface crosses the plane $b=0$ at $a=2$ and the plane $a=0$ at $b
\sqrt{x}=2$ for all values of $x$. For small $x$ we have $f_1>f_2$
and $f_2>f_1$ for $x>1$. On the surfaces shown on the figure by
red color (two curved triangles) there is a jump of functions
$f_i$ and we have the discontinuous  glass-glass transitions. In
Fig.~\ref{fig:fig4} we present  $f_1$ and $f_2$ as functions of
$b$ for $x=0.002$ and $a=1.9$, which are discontinuous when
passing through the red curved triangle. If the value of $x$ is
greater than $x_{02}$ we have the analogous behavior with
exchanged roles of $f_1$ and $f_2$. The overall picture is
symmetric in the variables $a$ and $b \sqrt{x}$. The plane of
variables ${a,b}$ for $x=1$
($x=\frac{S(k_2)}{S(k_1)}\frac{k_2^2}{k_1^2}$) is the boundary
between two glass phases: with $f_1>f_2$ and with $f_2>f_1$. This
transition is continuous. The behavior of the functions $f_i$ in
this case is shown in Fig.~\ref{fig:fig5}.

The surfaces of the discontinuous glass-glass transitions cross
the surface of liquid-glass transition and end in the point of the
$A_4$ singularities. This crossing line is one of the part of the
curved triangles, the other being the line of $A_3$ singularities.

The obtained qualitative picture is in good agreement with the
results of the MCT approach with Roger-Young approximation for the
direct correlation function $C(r)$ \cite{sperl1} or with
numerically obtained $C(r)$ for binary system \cite{das}. It is
also in good agreement with MD approach to the system \cite{das}.

To obtain quantitative phase diagram, one has to model the
dependence of $a,b,x$ on the physical variables and add the "tail"
to the structure factor.

The second part of such a program can be done in the following
way. Let us introduce in Eqs.~(\ref{Eq/Pg1/1:swal}) a
 small third term of the same structure. We have now instead of
 (\ref{Eq/Pg1/1:swal}) the following system:
\begin{eqnarray}
\frac{f_1}{1-f_1} &=& (a f_1+b f_2+\alpha  f_3)^2;\nonumber \\
\frac{f_2}{1-f_2} &=& x (a f_1+b f_2+ \alpha  f_3)^2;\nonumber \\
\frac{f_3}{1-f_3} &=& z (a f_1+b f_2+\alpha  f_3)^2.
\label{Eq/Pg4/1:swal}
\end{eqnarray}

 It is very important to note, that in this case the overall topology of the bifurcation surfaces
is changed entirely, we will obtain the singularity of the higher
order $A_6$, and the problem will be complicated to great extend
(see \cite{osob}). Again we have the multiplication of
singularities. Now there will be six points $A_6$ due to the
invariance of the Eq.(\ref{Eq/Pg4/1:swal}) connected with the
following transformations:

$$f_1 \to f'_2; f_2 \to f'_1; f_3 \to f'_3$$
$$a \to \sqrt{x'}b'; b \to \sqrt{x'}a'; \alpha  \to \sqrt{x'}\alpha '; x
\to 1/{x'}; z \to z'/{x'}
$$

$$f_1 \to f'_3; f_3 \to f'_1; f_2 \to f'_2$$
$$a \to \sqrt{z'}\alpha '; b \to \sqrt{z'}b'; \alpha  \to \sqrt{z'}a'; x
\to x'/{z'}; z \to 1/{z'}
$$

$$f_2 \to f'_3; f_3 \to f'_2; f_1 \to f'_1$$
$$a \to a'; b \to \alpha '; \alpha  \to b'; x \to z'; z
\to x'
$$

$$f_1 \to f'_2; f_2 \to f'_3; f_3 \to f'_1$$
$$a \sqrt{z} \to b'; b \sqrt{z} \to \alpha '; \alpha  \sqrt{z} \to a'; x
\to z'/{x'}; z \to 1/{x'}
$$

$$f_1 \to f'_3; f_2 \to f'_1; f_3 \to f'_2$$
$$a \sqrt{x} \to \alpha '; b \sqrt{x} \to a'; \alpha  \sqrt{x} \to b'; x
\to z'; z \to x'z'
$$

We will not try to investigate the overall picture of the
singularities surfaces, but will take into account the additional
terms as perturbation. One can imagine that it is possible to
consider the small shift of the two points $A_4$ obtained above
assuming that the coefficient $\alpha $ is small.

The function $f_3$ enters the equations with the multiplier
$\alpha $. This means that we can define $f_3$ in terms of
unperturbed values and use the third of the
 Eqs.(\ref{Eq/Pg4/1:swal}) as a definition of $f_3$.
In fact, as it can be seen from the consideration below, the
actual form of the perturbation term is not important. For
example, we can add the function $f_4$ with the small coefficient
$\beta $. Now we can use our previous result as the unperturbed
one in order to write the new functional $\hat\Psi (w,a,b,x)$
depending on the parameters $\alpha f_3$ and $\beta  f_4$, where
$\alpha,\beta  <<1$ and
$$f_3=\frac{zy_0^2}{1+zy_0^2},f_4=\frac{ty_0^2}{1+ty_0^2} .$$

In this case, for the $A_4$ singularity we have the conditions:
$\hat\Psi_w(w_0,...) = 0$, $\hat\Psi'_w(w_0,...) = 0$,
$\hat\Psi''_w(w_0,...) = 0$, $\hat\Psi'''_w(w_0,...) = 0$, Let us
expand the functional $\hat\Psi$ around the solution for $A_4$
singularity of $\Psi$, that is around $\Psi(y_0,a_0,b_0,x_0)$, up
to the first order in $\alpha,\beta  $. Now instead of the
relation (\ref {Eq/Pg1/3:swal}) we have:
\begin{equation}
w=\frac{aw^2}{1+w^2}+\frac{bxw^2}{1+xw^2}+ \alpha f_3 + \beta f_4;
\label{Eq/Pg5/1:swal}
\end{equation}
Within the proposed accuracy, it is irrelevant whether $f_3$ and
$f_4$ are written in terms of $w$ or $y_0$. So, we consider the
functional
\begin{equation}
\hat\Psi (y,a,b,x)= \gamma  + \Psi(y,a,b,x) \label{Eq/Pg5/2:swal}
\end{equation}
with $\Psi(y,a,b,x)$ given by  (\ref{Eq/Pg2/2:swal}) and $\gamma
=(\alpha f_3+ \beta f_4)/y_0$.

Let us note, that for $\gamma$ we can use functions taken at the
point $A_4$. In this approximate formulation, the topology of the
problem is not changed: we only have the shift of the singularity
point. The zero solution for $w$ also exists. Now let us put
$y=y_0+\gamma y_1$; $x=x_0+\gamma x_1$, $a=a_0+ \gamma a_1$,
$b=b_0+ \gamma  b_1$, expand the functional up to first order in
$\gamma $ and find the shifted point of singularity $A_4$. The
equations of zeroth order are fulfilled, and we obtain the system
of 4 linear equations for the variables $x_1$, $a_1$, $b_1$ and
$y_1$:
\begin{eqnarray}
\hat\Psi(y)&=&\Psi (0)+ \gamma  y_1 \Psi'_y(0)+\gamma
 \Psi_x'(0)x_1 + \nonumber \\
 &+&\gamma \Psi'_a a_1+\gamma  \Psi'_b(0)b_1+\gamma =0;\nonumber \\
\hat\Psi'(y)&=&\Psi_y'(0)+ \gamma  y_1 \Psi_{yy}''(0)+\gamma
\Psi''_{xy}(0)x_1+\nonumber \\
&+&\gamma  \Psi''_{ay}(0)a_1+\gamma \Psi''_{by}(0)b_1=0;\nonumber \\
\hat\Psi''(y)&=&\Psi_{yy}''(0)+ \gamma  y_1
\Psi_{yyy}'''(0)+\gamma \Psi'''_{xyy}(0)x_1+\nonumber \\
&+&\gamma\Psi'''_{ayy}(0)a_1+\gamma \Psi'''_{byy}(0)b_1=0;\nonumber \\
\hat\Psi'''(y)&=&\Psi_{yyy}'''(0)+ \gamma  y_1
\Psi_{yyyy}'''(0)+\gamma \Psi''''_{xyyy}(0)x_1+\nonumber \\
&+&\gamma\Psi''''_{ayyy}(0)a_1+\gamma \Psi''''_{byyy}(0)b_1=0.
\end{eqnarray}

This system has the form
\begin{eqnarray}
&&A_{11} x_1 + A_{12} a_1 +A_{13} b_1 +A_{14} y_1 = -1;\nonumber \\
&&A_{21} x_1 + A_{22} a_1 +A_{23} b_1 +A_{24} y_1 = 0;\nonumber \\
&&A_{31} x_1 + A_{32} a_1 +A_{33} b_1 +A_{34} y_1 = 0;\nonumber \\
&&A_{41} x_1 + A_{42} a_1 +A_{43} b_1 +A_{44} y_1 = 0;\nonumber
\end{eqnarray}
The coefficients $A_{ij}$ can be easily obtained by taking
derivatives of the equations (\ref{Eq/Pg2/2:swal} ) and
(\ref{Eq/Pg3/1:swal} )-(\ref{Eq/Pg3/3:swal} ) at the points $A_4$.
We have:
\begin{eqnarray}
&&A_{11} = b y + 3 y^4 -y^2; A_{12} = y + x y^3;\nonumber \\
&&A_{13} = xy + xy^3; A_{14} = 0;\nonumber \\
&&A_{21} = b + 8 y^3 - 2 y; A_{22} = 1 + 3 x y^2;\nonumber
\\
&&A_{23} = x + 3 x y^2; A_{24} = 0;\nonumber \\
&&A_{31} = 12y^2 -2; A_{32} = 6 x y;\nonumber \\
&&A_{33} = 6 x y; A_{34} = 0; \nonumber \\
&&A_{41} =0; A_{42} = 6 x;\nonumber \\
&&A_{43} = 6 x; A_{44} = - 24 x.\nonumber
\end{eqnarray}
Now  we can solve the system and obtain the values $x_1$, $a_1$,
$b_1$, and $y_1$. For example,
\begin{equation}
x_1=-\frac{3 }{y^2_0(1-3y^2_0)}. \label{Eq/Pg7/1:swal_1}
\end{equation}
This value is positive for the point $x_0=x_{01}, y_0=y_{01}$ and
is negative for $x_0=x_{02}, y_0=y_{02}$. So, the points $A_4$
move one towards another, and it is this result that we were going
to show: this fact makes our qualitative results closer to
numerical observations.

Let us recall, that in fact the topology of the bifurcation
surfaces has to be changed essentially if the higher degrees of
$f_i$ are taken into account.

The equations defining four variables $x_0, a_0, b_0, y_0$ at the
point $A_4$ for $\gamma=0$ are formally the same as the equations
for $A_4$ in the first order in $\gamma $. This means that we can
continue our iteration process up to $\gamma ^2$. Here $\gamma ^2$
denote the terms of the second order in $\alpha ,\beta $, which
has the form
$$\left(\frac {\alpha f_3+ \beta f_4}{y_0+\gamma y_1}\right)^2,$$
where $f_3$ and $f_4$ are written up to first order in $\gamma $.
This new shift is also of the form (\ref{Eq/Pg7/1:swal_1}) but
with $y_0$ changed for $y_0+\gamma y_1$. This means that for small
$\gamma $ the points $A_4$ continue to come nearer to one another.

It should be noted, that the obtained results do not contradict to
Ref.~\cite{sperl}, and one can show just in the same way as it is
done in \cite{GoSp02}, that in our model with $A_4$-singularities
the relaxation is logarithmic. The influence of multiplication of
singularities on the relaxation characteristics will be considered
in a separate paper.

To conclude, the SS system is considered in the frame of Mode
Coupling Theory approach. An approximation for the structure
factor is used that emphasizes the quasibinary character of the
system. The approximation enables to obtain in an analytic way the
solution of the MCT equations for the infinite time limit and
focus on the symmetry of the problem. The bifurcation singularity
is $A_4$, and, taking into account the symmetry, gives the "double
swallow tail". The qualitative phase diagram is constructed in the
variables
$$x=\frac{S(k_2)}{S(k_1)}\frac{k_2^2}{k_1^2},
a=\frac{S(k_1)k_1}{8 \pi^2 \rho }, b=\frac{S(k_2)k_2}{8 \pi^2 \rho
} $$ where $k_1$ and $k_2$ are the locations of two maxima of the
structure factor. Besides the liquid-glass transition, the phase
diagram includes a plane $x=1$ of the continuous glass-glass
transitions, and the discontinuous glass-glass transitions occur
for $x<x_{01}$ and $x>x_{02}$.

\bigskip

\section{acknowledgments}
We are greateful to S. M. Stishov, and V. V. Brazhkin, for
stimulating discussions. The work was supported in part by the
Russian Foundation for Basic Research (Grants No 11-02-00341,
13-02-12008, 13-02-91177, and 13-02-00579).

\end{document}